\documentclass[twocolumn,trackchanges]{aastex631}

\newcommand\rp{\emph{r}-process}
\newcommand\rpe{\emph{r}-process-enhanced}

\usepackage{enumerate}
\usepackage{graphicx}
\usepackage{amsmath}
\usepackage{float}
\usepackage{xcolor}
\usepackage{hyperref}
\hypersetup{
    bookmarks=true,			
    unicode=false,			
    pdftoolbar=true,			
    pdfmenubar=true,			
    pdffitwindow=true,			
    pdfstartview={FitH},		
    pdftitle={\mbox{HD 222925}},
    pdfauthor={eholmbeck},	
    pdfsubject={Astrophysics},	
    pdfcreator={dvipdf},		
    pdfproducer={dvipdf},		
    pdfkeywords={r-process, stars, nucleosynthesis}, 
    pdfnewwindow=true,		
    colorlinks=true,			
    linkcolor=magenta,		
    citecolor=teal,			
    filecolor=magenta,			
    urlcolor=cyan,			
    breaklinks=true,
    linktocpage
}

\received{14 October 2022}
\submitjournal{\apj}

\shorttitle{Astrophysical and Nuclear Conditions of \emph{r}-process Sites}
\shortauthors{Holmbeck et al.}

\begin{document}

\title{\mbox{HD 222925}: a New Opportunity to Explore the Astrophysical and Nuclear Conditions of \emph{r}-process Sites}

\correspondingauthor{Erika M.\ Holmbeck}
\email{eholmbeck@carnegiescience.edu}

\author{Erika M.\ Holmbeck}
\affiliation{Observatories of the Carnegie Institution for Science, 813 Santa Barbara St., Pasadena, CA 91101, USA}
\affiliation{Joint Institute for Nuclear Astrophysics -- Center for the Evolution of the Elements (JINA-CEE), USA}
\affiliation{Hubble Fellow}

\author{Rebecca Surman}
\affiliation{Department of Physics, University of Notre Dame, Notre Dame, IN 45565, USA}
\affiliation{Joint Institute for Nuclear Astrophysics -- Center for the Evolution of the Elements (JINA-CEE), USA}

\author{Ian U.\ Roederer}
\affiliation{Department of Astronomy, University of Michigan, 1085 S.\ University Ave., Ann Arbor, MI 48109, USA}
\affiliation{Joint Institute for Nuclear Astrophysics -- Center for the Evolution of the Elements (JINA-CEE), USA}

\author{G.\ C.\ McLaughlin}
\affiliation{Department of Physics, North Carolina State University, Raleigh, NC 27695, USA}
\affiliation{Joint Institute for Nuclear Astrophysics -- Center for the Evolution of the Elements (JINA-CEE), USA}

\author{Anna Frebel}
\affiliation{Department of Physics and Kavli Institute for Astrophysics and Space Research, Massachusetts Institute of Technology, Cambridge, MA 02139, USA}
\affiliation{Joint Institute for Nuclear Astrophysics -- Center for the Evolution of the Elements (JINA-CEE), USA}

\begin{abstract}
With the most trans-iron elements detected of any star outside the Solar System, \mbox{HD 222925} represents the most complete chemical inventory among \emph{r}-process-enhanced, metal-poor stars. 
While the abundance pattern of the heaviest elements identified in \mbox{HD 222925} agrees with the scaled Solar \emph{r}-process residuals, as is characteristic of its \rpe\ classification, the newly measured lighter \emph{r}-process elements display marked differences from their Solar counterparts.
In this work, we explore which single astrophysical site (if any) produced the entire range of elements ($34\leq Z\leq 90$) present in \mbox{HD 222925}.
We find that the abundance pattern of lighter \emph{r}-process elements newly identified in \mbox{HD 222925} presents a challenge for our existing nucleosynthesis models to reproduce.
The most likely astrophysical explanation for the elemental pattern of \mbox{HD 222925} is that its light \emph{r}-process elements were created in rapidly expanding ejecta (e.g., from shocked, dynamical ejecta of compact object merger binaries).
However, we find that the light \emph{r}-process-element pattern can also be successfully reproduced by employing different nuclear mass models, indicating a need for a fresh investigation of nuclear input data for elements with $46 \lesssim Z \lesssim 52$ by experimental methods.
Either way, the new elemental abundance pattern of \mbox{HD 222925}---particularly the abundances obtained from space-based, ultraviolet (UV) data---call for a deeper understanding of both astrophysical \emph{r}-process sites and nuclear data.
Similar UV observations of additional \rpe\ stars will be required to determine whether the elemental abundance pattern of \mbox{HD 222925} is indeed a canonical template for the \emph{r}-process at low metallicity. 
\end{abstract}



\section{Introduction}
\label{sec:intro}

Recently, \citet{Roederer2022} derived chemical abundances of the bright, metal-poor (${\rm[Fe/H]} = -1.46$) star \mbox{HD 222925} in and around the second \rp\ peak from UV observations.
These new measurements place \mbox{HD 222925} as the star with the most elements measured in its spectrum, second only to the Sun.
Totaling 42 trans-iron elements ($Z>30$), \mbox{HD 222925} is only missing a confident abundance determination for eight of the neutron-capture elements that have stable or long-lived isotopes.
Like many \rp-rich stars, the main \rp\ pattern ($Z\geq 56$) of \mbox{HD 222925} shows stunning agreement with the Solar \rp\ residuals.
However, its pattern diverges from the Solar pattern among the light \rp\ elements ($34\leq Z \leq 52$, see Figure \ref{fig:sprocess}).
Hints of this discrepancy were seen with ground-based, optical observations \citep{Roederer2018} and are now more apparent with spaced-based, UV observations.
Comparisons to other \rpe, metal-poor stars indicate that the abundance pattern of \mbox{HD 222925} is not atypical for its type; it appears to differ from other \rpe\ stars only in the number of elements measured.
Having the most complete abundance pattern of any star outside of the Solar system, along with a low metallicity, \mbox{HD 222925} could be a new, better, standard by which to compare \rp\ nucleosynthesis simulations.
However, the divergence from the Solar pattern among the lighter \rp\ elements raises new questions about the \rp\ site.
Is the elemental abundance pattern of \mbox{HD 222925} fully representative of other \emph{r}-process-enhanced stars at low metallicity, and if so, what does this mean for the \emph{r}-process site that enriched the gas from which they formed?

Nucleosynthesis calculations within the past few decades have expanded from comparing their simulation yields to only the Solar abundance pattern to including in their comparisons the abundance patterns of \rp-rich, metal-poor stars \citep[e.g.,][]{Wanajo2013,Hansen2014,Vassh2020}.
The Solar system abundances are still the most complete observational information we have for the \rp, since they can be split into contributions on the \emph{isotopic} level.
In contrast, abundances derived from other stellar spectra are necessarily elemental (except for a handful of cases in which approximate isotopic ratios can be determined).
Unlike the Sun, however, the low metallicity of \rpe\ stars suggests fewer progenitor sources of heavy elements and a higher likelihood that a single simulation output accurately represents the \rp\ source that originally enriched the star's prenatal gas cloud.
In this work, we expand on these recent efforts in nucleosynthesis studies by using \mbox{HD 222925} to reconstruct astrophysical conditions of (potentially canonical) \rp\ sites at low metallicity.

There is a long tradition in \rp\ studies of approximating the combination of astrophysical conditions that best reproduces the Solar \rp\ abundances \citep{Kratz1993,Freiburghaus1999}.
In the same spirit, \citet{Holmbeck2019a} introduced the ``Actinide-Dilution with Matching" (ADM) model to characterize the physical differences between \rp\ sites as a function of elemental abundance differences in \emph{r}-process-enhanced, metal-poor stars.
This model builds a distribution of astrophysical conditions (e.g., ejecta entropy or initial electron fraction) by simultaneously matching key elemental abundance ratios derived from observed stellar spectra through randomly sampling nucleosynthesis network abundance outputs corresponding to different astrophysical conditions.
Specifically, this model can explain the star-to-star differences in actinide abundances, finding that a moderate increase of material ejected at very low initial electron fractions ($Y_e$) could sufficiently explain all observed levels of actinide enhancements.

In this work, we use the ADM approach to explore possible combinations of astrophysical ejecta that could potentially be responsible for the \mbox{HD 222925} abundance pattern, including the distinctly non-Solar light \emph{r}-process element abundances.
In doing do, we attempt to reconstruct the progenitor astrophysical conditions which gave rise to the heavy-element pattern of \mbox{HD 222925}.
If \mbox{HD 222925} is to be the new standard \emph{r}-process abundance pattern template, we will have effectively reconstructed the canonical \emph{r}-process site responsible for the observational signatures in metal-poor stars.
First, in Section \ref{sec:trends} we examine notable features of \mbox{HD 222925} that the ideal site reconstruction will reproduce.
Next, in Section \ref{sec:site} and \ref{sec:data} we apply the ADM method to \mbox{HD 222925} and examine the suitability of its pattern as representative of the canonical \emph{r}-process site under different astrophysical and nuclear conditions.
Finally, we comment on what the results in Section \ref{sec:site} and \ref{sec:data} reveal about the uniquely complete abundance pattern of \mbox{HD 222925} and the \rp\ that enriched the gas from which it formed.

\section{Key Elemental Trends of \mbox{HD 222925}}
\label{sec:trends}

\begin{figure}[t]
	\centering
	\includegraphics[width=\columnwidth]{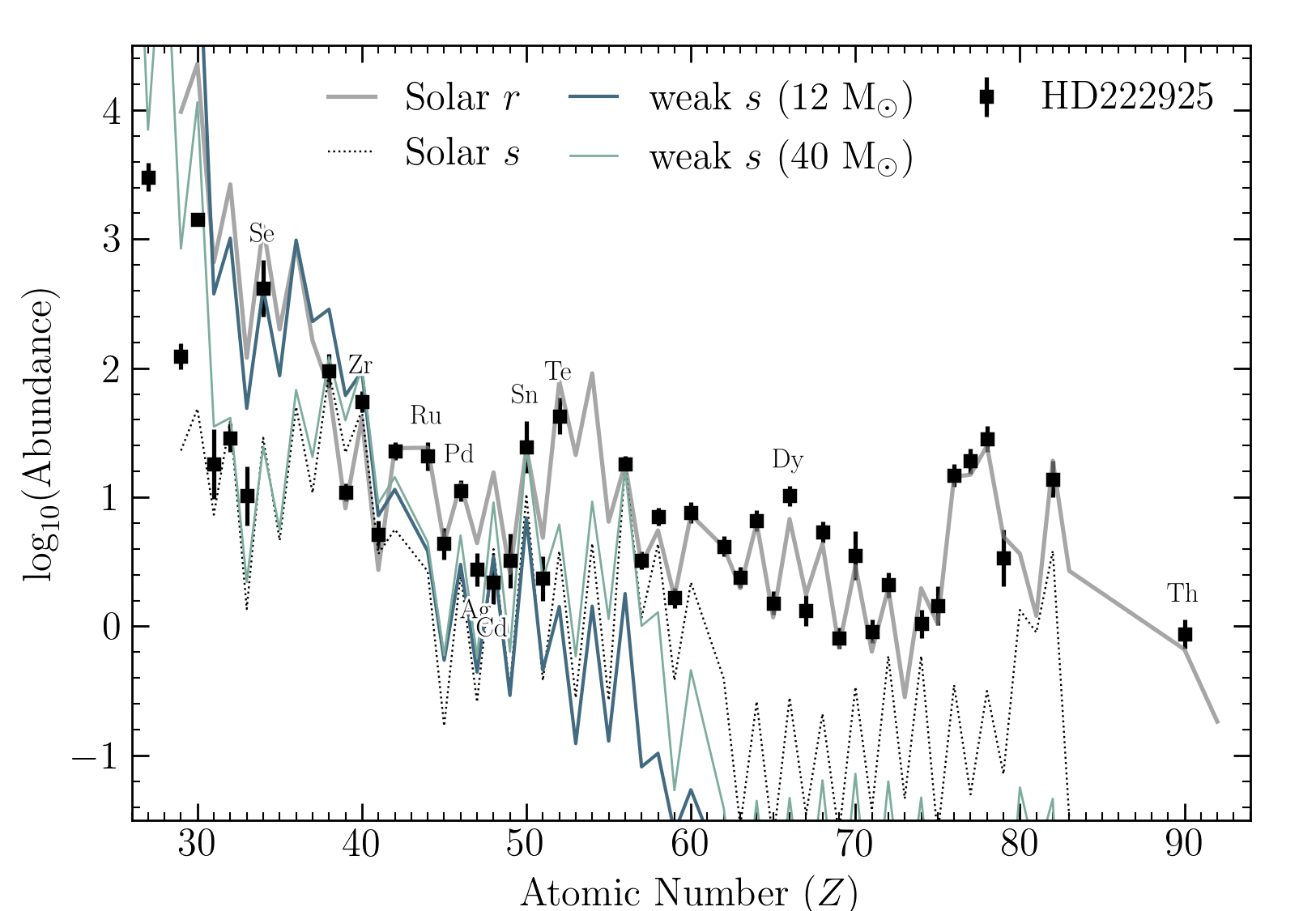}
	\caption{Abundance patterns of \mbox{HD 222925} (black points) compared to the Solar \emph{r} and \emph{s}-process components (gray lines) and weak \emph{s}-process models from \citet{Banerjee2018} (green lines). The Solar neutron-capture patterns and the 40-M$_\odot$ \emph{s}-process model are scaled to Ba, while the 12-M$_\odot$ model is scaled to Se.
	}
	\label{fig:sprocess}
	\end{figure}

In our previous application of the ADM model \citep{Holmbeck2019a}, only three elements were used: Zr, Dy, and Th, chosen in order to gauge the relative amounts of light (Zr) and actinide (Th) material compared to the main (Dy) \rp.
In that work, we were not as concerned with specific matches to the entire detailed abundance pattern, and rather argued that too many elemental constraints may lead to an artificially constrained result, especially considering the state of nuclear physics uncertainties and their effect on \rp\ calculations.
However, \mbox{HD 222925} displays specific abundance features, unique in their UV observations, with which we seek to supplement and challenge our Zr-Dy-Th-based model.
Here we comment on those elements which distinguish this star from the Solar \emph{r}-process residuals and how those particular elements will guide our ADM model choices.
For reference, Figure \ref{fig:sprocess} shows the abundance pattern of \mbox{HD 222925} compared to the Solar \emph{s}- and \emph{r}-process abundances, scaled to the abundance of Ba.

\subsection{Se and the \emph{s}-process}

In the lightest portion of its heavy-element abundance pattern, \mbox{HD 222925}'s high Se abundance relative to its neighbors stands out.
This element is dominated by the \emph{r}-process in the Solar system; however, it is worth noting that Se can be produced by a low-metallicity, weak \emph{s}-process.
We show in Figure \ref{fig:sprocess} two weak \emph{s}-process models: progenitors of mass 12 and 40~M$_\odot$ with $[\textrm{Fe/H}]=-3$ and $[\textrm{CNO/H}]=0$ from \citet{Banerjee2018}.
Although a limited \emph{s}-process could account for the high abundances of elements such as Se and Sn, such weak \emph{s}-process models tend to overproduce elements lighter than Se (e.g., Zn through As, $30\leq Z\leq 33$), as shown by the dark green (12 M$_\odot$) line.
Attempts to mitigate the overproduction of the lighter elements with different \emph{s}-process models underproduces Se by an order of magnitude at the cost of overproducing Ba, shown by the light green (40 M$_\odot$) line.
Similarly, models that can reproduce the high Sn are an otherwise poor match to the other elements in \mbox{HD 222925}.
We support the claim made in \citet{Roederer2022} that \mbox{HD 222925} is not so significantly \emph{s}-process enriched as to account for a significant production of the lighter elements like Se and Sn and attribute the majority of their origin to the \emph{r}-process.

We note that the abundance of Se was derived from one absorption feature in the UV; however, the absorption feature is strong, well-isolated, and not suspected to greatly suffer from hyperfine-splitting effects, isotopic shifts, or assumptions about non-local thermodynamic equilibrium \citep{Roederer2012,Peterson2020}.
Our initial ADM model will not use Se as a constraint (since this model did not sample enough parameter space for our \emph{r}-process models to robustly produce it), but we will enforce Se production in a more complex model (see Section~\ref{sec:cold}) as a noteworthy feature in need of explanation.

\subsection{Zr to Te}

	\begin{figure*}[t]
	\centering
	\includegraphics[height=2.6in]{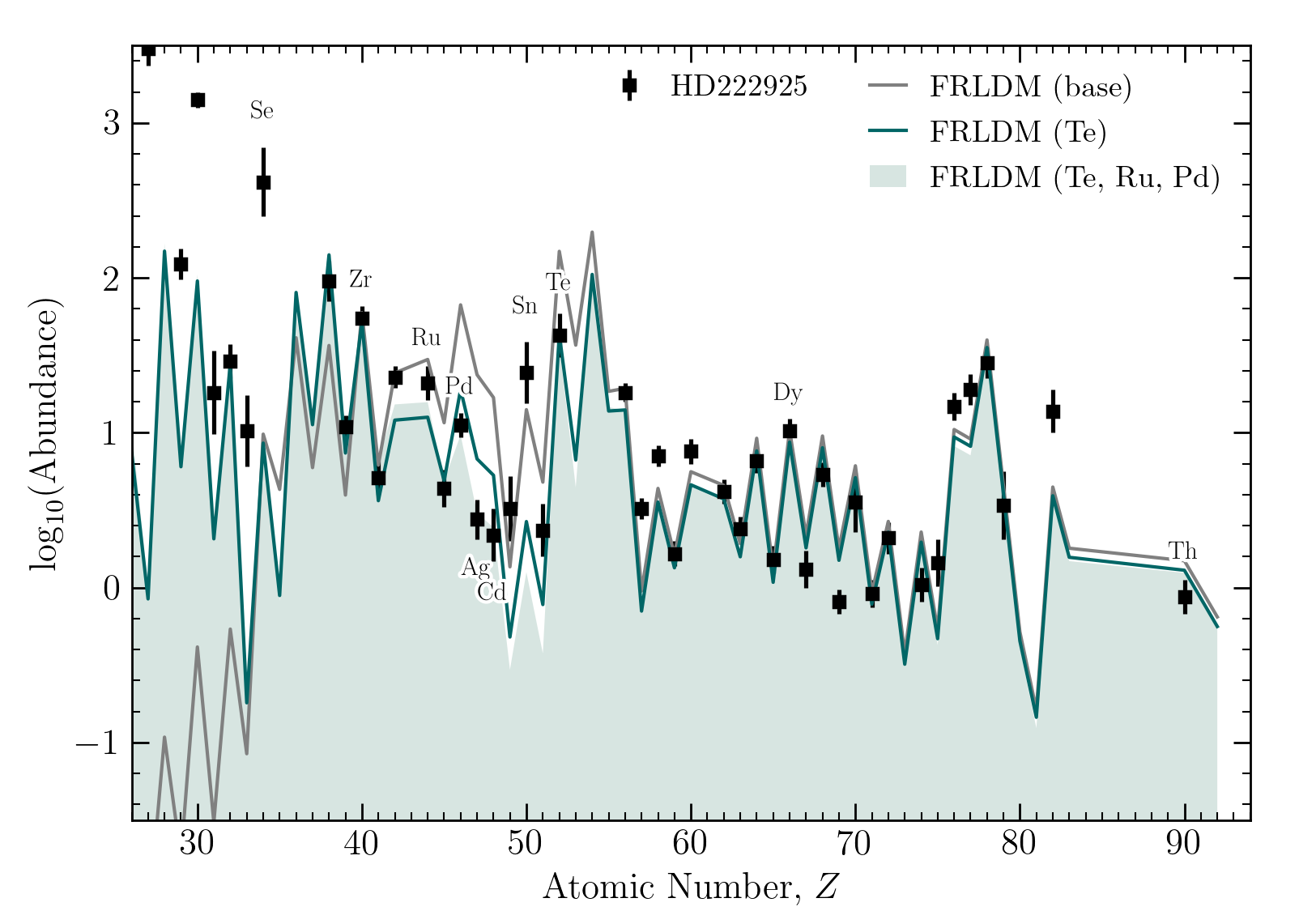}
	\includegraphics[height=2.6in]{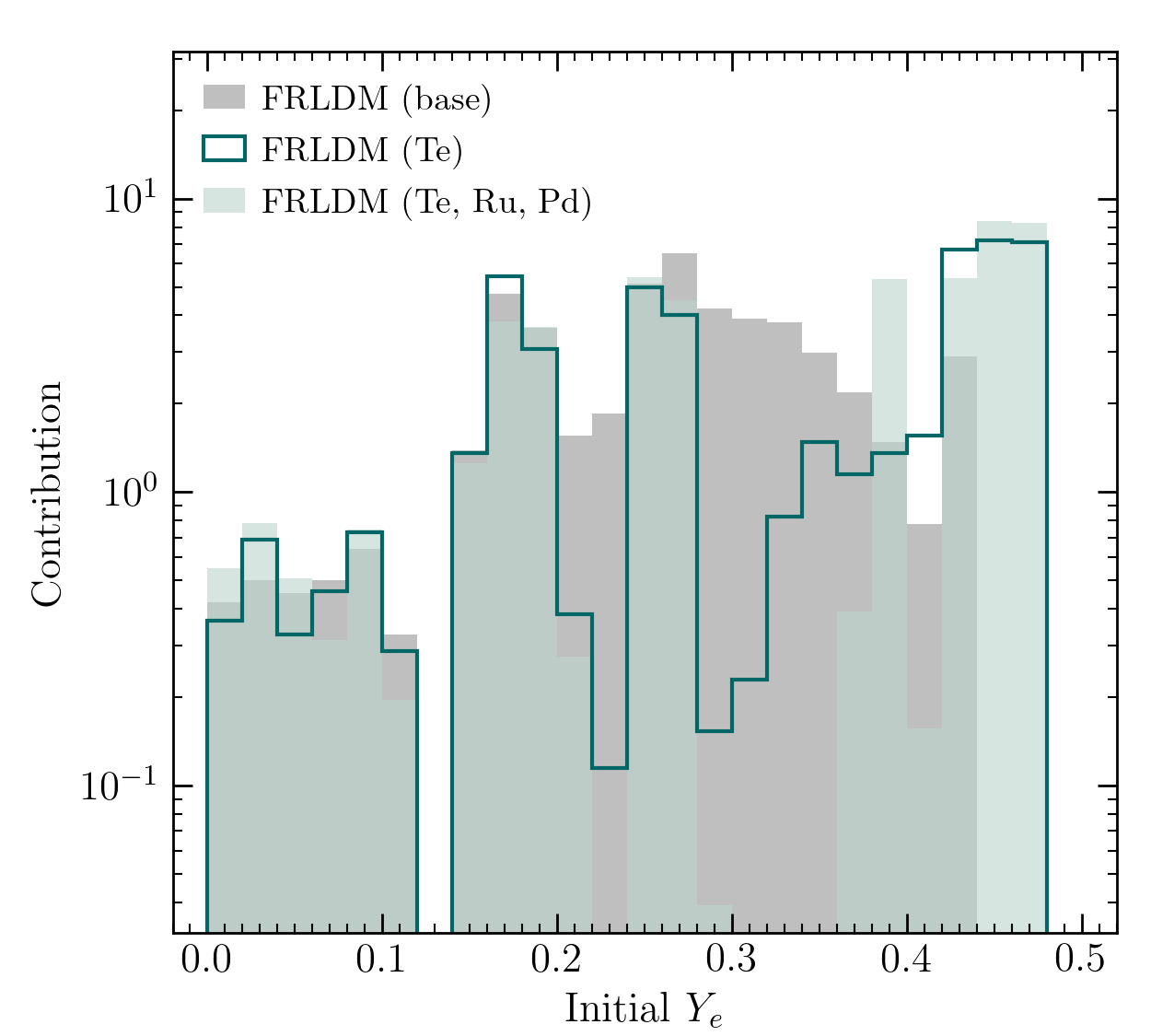}
	\caption{Combined abundance patterns (left) and $Y_e$ distributions of ejecta (right) predicted by the model when two (gray), three (teal), and five (green filled) elemental ratios are used. The individual nucleosynthesis calculations use the FRLDM fission fragment distribution for actinide-producing conditions ($Y_e\lesssim 0.25$).}
	\label{fig:s040-korobkinrh_frldm}
	\end{figure*}

Perhaps the most intriguing feature in the abundance pattern of \mbox{HD 222925} is the rapid departure from the Solar \rp\ pattern for the $40\leq Z\leq 48$ elements, followed by an agreement with Solar at In and Sn ($Z=49$ and 50, respectively) and lower abundances again for the next two elements, Sb and Te.
Also note that the Solar \emph{r}-process component has not been scaled; the \emph{r}-process enhancement of \mbox{HD 222925} is so significant that its absolute abundances are on par with that of the Solar system \citep{Roederer2018}.
The abundances for the four elements with $Z=48$--51 were derived from only one spectral feature each, and thus large uncertainties accompany their adopted abundance values.
While we do not invoke their uncertainties to dismiss these measurements, we place less importance on explaining their trends than the other elements in this region.
In particular, we will focus on the relationship between Zr, Ru, Pd, Ag, and Te, as well as their overall co-production with the main \rp\ pattern ($Z\geq 56$).

\subsection{Ba to Th}

In order to investigate whether one \emph{r}-process source can synthesize the entire range of abundances in HD22295, we will also require our model to produce lanthanide and actinide abundances (represented by Dy and Th, respectively, as per the original ADM model) commensurate with those derived for the star.
We focus less on fitting the element-by-element details of this main \emph{r}-process component since nuclear uncertainties dominate in this region and might lead to overfitting.
As in \citet{Holmbeck2019a}, we assume the Th abundance at 1 Gyr post-event was 0.2~dex greater than the modern-day Th abundance derived for the star in order to account for roughly 10 Gyr of radioactive decay, commensurate with its low metallicity.

\section{\emph{r}-process sites}
\label{sec:site}

\subsection{The base model}
\label{sec:base}
	
To reproduce the abundance trends of \mbox{HD 222925}, we first run a calculation based on the ADM model of \citet{Holmbeck2019a}.
In this model, we randomly sample nucleosynthesis simulation output for a set of initial astrophysical conditions (entropy and density) with different starting compositions (calculated from nuclear statistical equilibrium at 10 GK with varying initial $Y_e$).
This variation provides a conceptual insight for the distribution of \rp\ ejecta from an astrophysical site that could explain the observed abundance trends, and the $Y_e$ distribution constructed by previous applications of the ADM model has demonstrated agreement with those from robust hydrodynamical simulations.
We start with our base, two-ratio model that uses the Zr/Dy and Th/Dy abundance ratios and will only add complexities later as needed to explain the data.

First we test a combination of ejecta with two distinct entropies.
We run a suite of nucleosynthesis calculations that follow one of two of the same astrophysical trajectories (temperature and density evolution over time), but vary the initial composition.
All nucleosynthesis simulations are run with the nuclear reaction network code Portable Routines for Integrated nucleoSynthesis Modeling \cite[PRISM;][]{Sprouse2020} and use nuclear masses from the Finite Range Droplet Model \citep[FRDM2012;][]{Moller2012,Moller2016} and $\beta$-decay rates from \citet{Moller2019}.
Laboratory-measured nuclear masses from the 2020 compilation of the Atomic Mass Evaluation \citep[AME2020;][]{Wang2021} and decay rates from the 2020 NUBASE evaluation \citep{NuBase2020} are used instead of the theoretical values wherever available.
For astrophysical conditions, we choose a very low-entropy ($s\sim 8~k_b/{\rm baryon}$) trajectory characteristic of dynamical ejecta from a neutron star merger and a moderately low entropy ($s\sim40~k_b/{\rm baryon}$) trajectory representing disk-wind-like conditions.
In the lower entropy case, we vary the $Y_e$ up to 0.25.
The $Y_e$ for the wind-like trajectory is allowed to vary between 0.25 and 0.5.
A separation at $Y_e=0.25$ was chosen since, for these entropies, this $Y_e$ is approximately the transition point at which lanthanides and actinides are synthesized in the \rp.
We use a broad fission fragment distribution based on the Finite Range Liquid Droplet Model (FRLDM) of \citet{Moller2015,Mumpower2020}, as employed by \citet{Vassh2020}.
Note that at the entropies considered here, only the most neutron-rich conditions ($Y_e \lesssim 0.20$) will achieve fission; in other words, only the ``dynamical-like" conditions will allow fission cycling for this subset of nucleosynthesis simulations.

The ADM model randomly samples output from the above trajectories in their allowed $Y_e$ ranges and finds the combination of 15 samples whose total abundance ratios are within given tolerances (typically about 0.1 dex).
All ratios must be simultaneously be within acceptable tolerances for the 15 samples to be considered a match.
For the base model, we use the abundance ratios of Zr/Dy and Th/Dy, as in the original ADM application.
The gray line in Figure \ref{fig:s040-korobkinrh_frldm} shows the result of the ADM model compared to \mbox{HD 222925} using these two constraints.
As constrained by the model, the lighter \emph{r}-process elements (Zr), lanthanides (Dy), and actinides (Th) are overall a decent match to the stellar data, and the $Y_e$ distribution is overall continuous, agreeing roughly with previous applications \citep{Holmbeck2019a}.\footnote{The gap in the distribution at $Y_e\approx 0.12$ is due to an (unfavorable) excess of actinide abundance, as described in \citet{Holmbeck2019a} \citep[see also][]{Eichler2019}.}
This baseline ADM application finds remarkable success with effectively only three elements; however, there are a few regions in which the ADM result departs from observational data.
For example, the region between Pd and Te is significantly overproduced by the base model, suggesting that the $Y_e$ distribution in Figure~\ref{fig:s040-korobkinrh_frldm} is not representative of the ``canonical" \rp\ site that produced the heavy elements in \mbox{HD 222925}.
The overproduction in this region was not noteworthy in \citet{Holmbeck2019a} since none of the stars considered in that sample had abundance derivations for the elements in this region.
Now, however, our model must account for the new data in this region.
It should also be noted that when this exercise was repeated with the Solar \emph{r}-process abundances, the reconstructed $Y_e$ distribution was similarly continuous as in the ``base" case in Figure~\ref{fig:s040-korobkinrh_frldm}.
Note again that Se also does not match with our ADM model, and none of the $Y_e$ variations for the two entropy cases produces sufficiently high Se in order to obtain such a match.
With the current model parameters, it is possible that Se could be produced at $Y_e > 0.5$ by, e.g., the $\nu p$-process or a very weak \emph{r}-process \citep{Psaltis2022}.
Se aside, it is clear we can invoke more constraints to obtain a better fit to the \mbox{HD 222925} abundance pattern, especially in the light \emph{r}-process-element region: $44\lesssim Z \lesssim 52$.

\subsection{Adding more constraints}

Since the region from Pd to Te is overproduced by the baseline ADM model, we expand the model by adding more constraints in addition to the original Zr, Dy, and Th in order to find the conditions of the canonical site responsible for the \rp\ pattern of \mbox{HD 222925}.
For this exercise, we add Te, Ru, and Pd as additional constraints to the model.
In practice, we use the abundance ratios Zr/Te, Zr/Pd, Pd/Ru, Te/Dy, and Th/Dy from \mbox{HD 222925} in order to capture the downward trend of the elements between Zr and Te.
There are regions in the heavy \emph{r}-process-element ($Z\geq 56$) abundance pattern that are also a less-than-ideal fit: for example, the underproduction of elements to the right of the second \emph{r}-process peak at $57\leq Z\leq 63$.
In addition, the overall shape and location of the third \emph{r}-process peak could also be affected by late-time neutron captures \citep{Surman2001,Eichler2015}.
We do not add more constraints to this portion of the abundance pattern, however, since we use this work to focus on the light \emph{r}-process region where the abundance pattern of \mbox{HD 222925} significantly diverges from the Solar \emph{r}-process component.

The green line and shaded region in the left panel of Figure \ref{fig:s040-korobkinrh_frldm} shows the total abundance results when more constraints are added to the base set: first only Te (in the form of Zr/Te), then with the addition of Ru and Pd, respectively.
Note that the abundances of Zr, Dy, and Th (the ``base" set) are still being used by the model.
Even adding the additional Te ratio (solid green line) begins to capture the downward trend of Pd, Ag, and Cd.
(For the Solar case, adding Te did not significantly change the $Y_e$ distribution.)
Adding even more constraints (shaded region) lowers this region even further for near-perfect agreement with Ag and Cd.
We cannot replicate the high abundances of In, Sn, and Sb, and many of the elements between Ba and Gd are similarly still underproduced.
However, seeing as how the uncertainties on the abundance derivations of In, Sn, and Sb are large, we count these fits as overall successes.

The right panel of Figure~\ref{fig:s040-korobkinrh_frldm} shows a histogram of the $Y_e$ combinations that contributed to the overall matching abundance ratios.
Unlike the mostly continuous $Y_e$ distribution in the base model, requiring the model to fit more elements leads to a more constrained $Y_e$ distribution.
With the addition of just one more element (Te), the $Y_e$ distribution begins to disfavor $Y_e\approx 0.25$ and $0.30$ (compare the gray shaded histogram with the green outlined one).
Adding two more elements for matching yields an even more constrained distribution that shows preference for four specific values of initial $Y_e$: $\sim$0.09, $\sim$0.17, $\sim$0.27, and $\sim$0.45.
(Recall that for $0\leq Y_e \leq 0.25$, the $s\sim8~k_b/{\rm baryon}$ were used, and $s\sim40~k_b/{\rm baryon}$ for $Y_e\geq 0.25$.)

	\begin{figure}[t]
	\centering
	\includegraphics[width=\columnwidth]{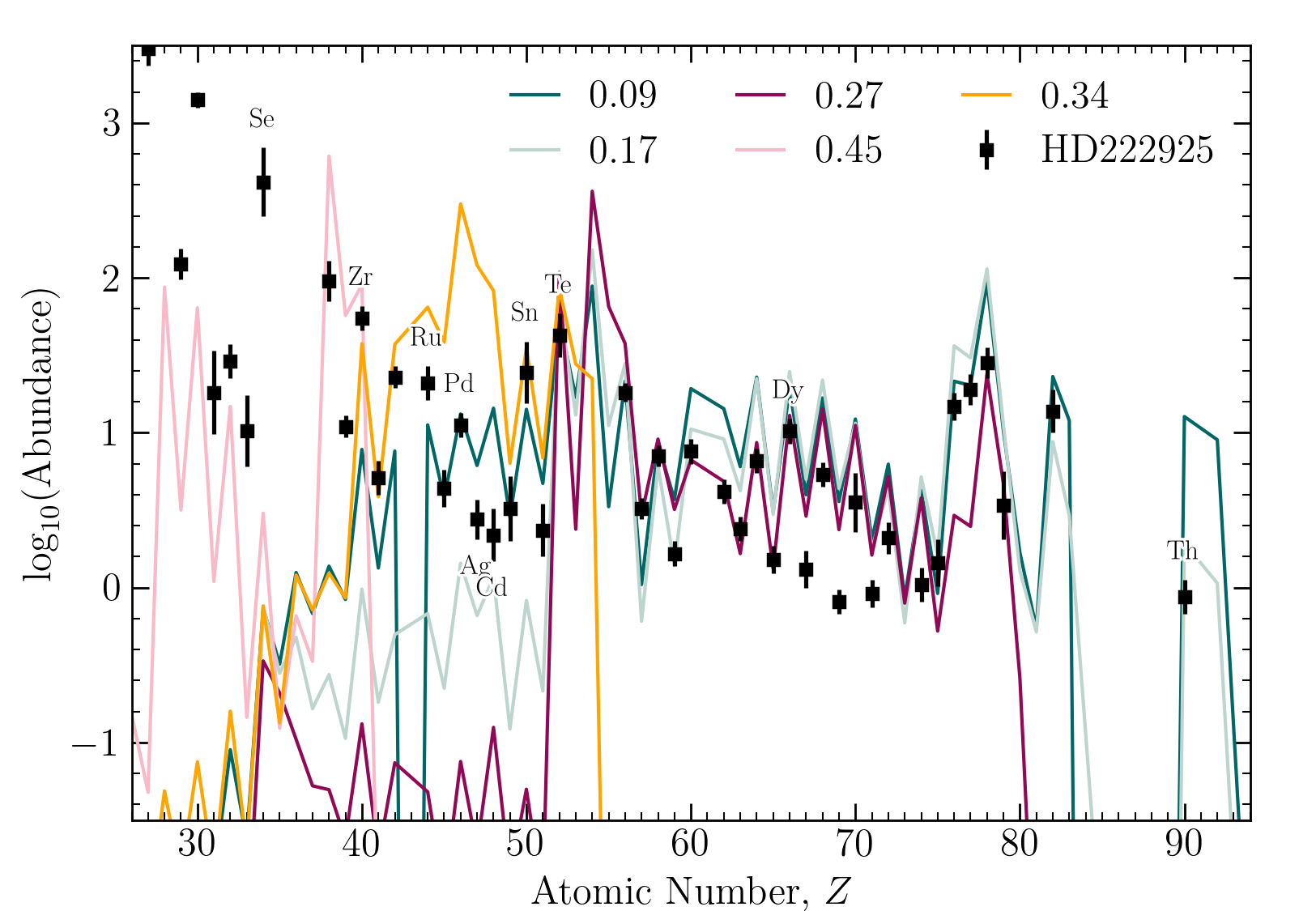}
	\caption{Abundance patterns corresponding to individual initial $Y_e$ values. The lowest values ($Y_e = 0.09$ and 0.17) use very low entropy ``dynamical" conditions with FRLDM fission yields, and the rest ($Y_e=0.27$, 0.45, and 0.34) use the moderately low entropy ``wind" conditions.
	}
	\label{fig:s040-korobkinrh_ye}
	\end{figure}

Figure~\ref{fig:s040-korobkinrh_ye} illustrates why these $Y_e$ values are preferred; the fission fragments from low-entropy, low-$Y_e$ conditions contribute most of the Pd abundance, and some can produce sufficiently high lanthanides without overproducing the actinides or Pd by fission deposition; compare the two green lines for $Y_e=0.09$ and 0.17.
Higher-$Y_e$ conditions contribute most of the Zr and Te abundance, as shown by the pink lines for $Y_e=0.27$ and 0.45.
Figure~\ref{fig:s040-korobkinrh_ye} also begins to show why Se cannot be used as a constraint for these initial models; the entropies and $Y_e$ values that the model samples from is simply not sufficient to produce Se.
We will return to Se in the next section by considering an astrophysical trajectory that does produce it.

Intriguingly, Zr and Te production appear decoupled from one another; there is a specific component that produces Zr and essentially none of the other elements ($Y_e\approx 0.45$), while Te can be produced by a variety of initial $Y_e$ values, but notably at $Y_e\approx 0.27$.
This separation contradicts recent evidence in metal-poor stars that Zr and Te are in fact correlated \citep{Roederer2022b}.
Moreover, this separation occurs at $Y_e\approx 0.34$ (orange line).
At these values of $Y_e$, Zr and Te are coproduced, but this model disfavors such coproduction as an explanation for \mbox{HD 222925} due to the extreme overabundance of elements like Pd that also occurs at this $Y_e$.

In principle the preference for certain conditions (here, entropy and $Y_e$) is unsurprising; additional distinct conditions are necessary to fit more abundance features in the same way early \rp\ studies necessarily invoked separate astrophysical conditions to explain the three \rp\ peaks in the Solar abundance pattern \citep[e.g.,][]{Kratz1993}.
However, in the case of the Sun, it can be argued that the need for a variety of conditions is evidence for multiple \rp\ sites contributing to the heavy-element abundances in the Solar system.
If the progenitor site of \mbox{HD 222925} is to be canonical of progenitors of \rpe\ stars, then the logical assumption is that the $Y_e$ distribution in Figure~\ref{fig:s040-korobkinrh_frldm} describes a single \rp\ site, not multiple.
With this in mind, we discuss four ways to interpret the $Y_e$ distribution for the most constrained case in Figure \ref{fig:s040-korobkinrh_frldm} (green shaded region):
	\begin{enumerate}
	\item At face value: the ejecta responsible for the \emph{r}-process enrichment of \mbox{HD 222925} did not have a continuous distribution of conditions, and specifically $s \approx 40$ with $Y_e \approx 0.34$ conditions were not or could not be achieved in the ejecta.
	\item As evidence for at least two astrophysical sites: one in which conditions produce ejecta with $Y_e\gtrsim 0.35$ and another with $Y_e\lesssim 0.30$.
	\item As a challenge to one of the basic assumptions of our ADM model: that we can represent the entirely of the ejecta with only two values of initial entropy.
	\item As a call for new laboratory studies of the nuclear data that shapes the abundance pattern features between the first and second \emph{r}-process peaks.
	\end{enumerate}

If the first case is correct, a precise balance in ejecta $Y_e$ would be needed in order to reproduce the main features of the abundance pattern of \mbox{HD 222925}, with sufficient $Y_e\approx 0.40$ for Zr, and no ejecta with $Y_e\approx 0.34$.
While possible, such a constraint seems unlikely; not only do many detailed NSM simulations easily predict ejected material with 0.34, but the distributions of $Y_e$ in the simulated ejecta are typically smooth in general with no such abrupt gaps, similar to the baseline distribution shown in Figure~\ref{fig:s040-korobkinrh_frldm} in gray.

	\begin{figure*}[t]
	\centering
	\includegraphics[height=2.6in]{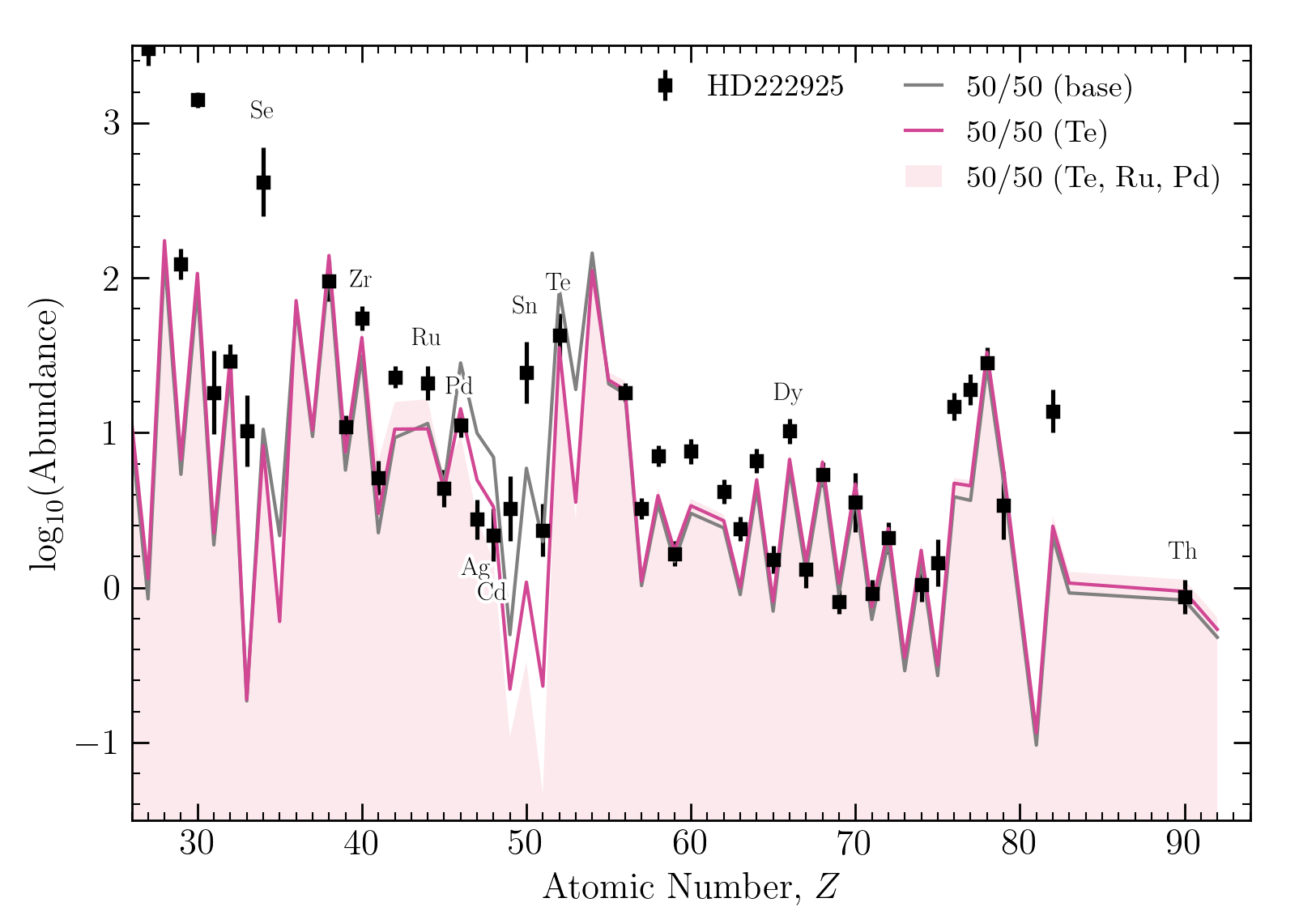}
	\includegraphics[height=2.6in]{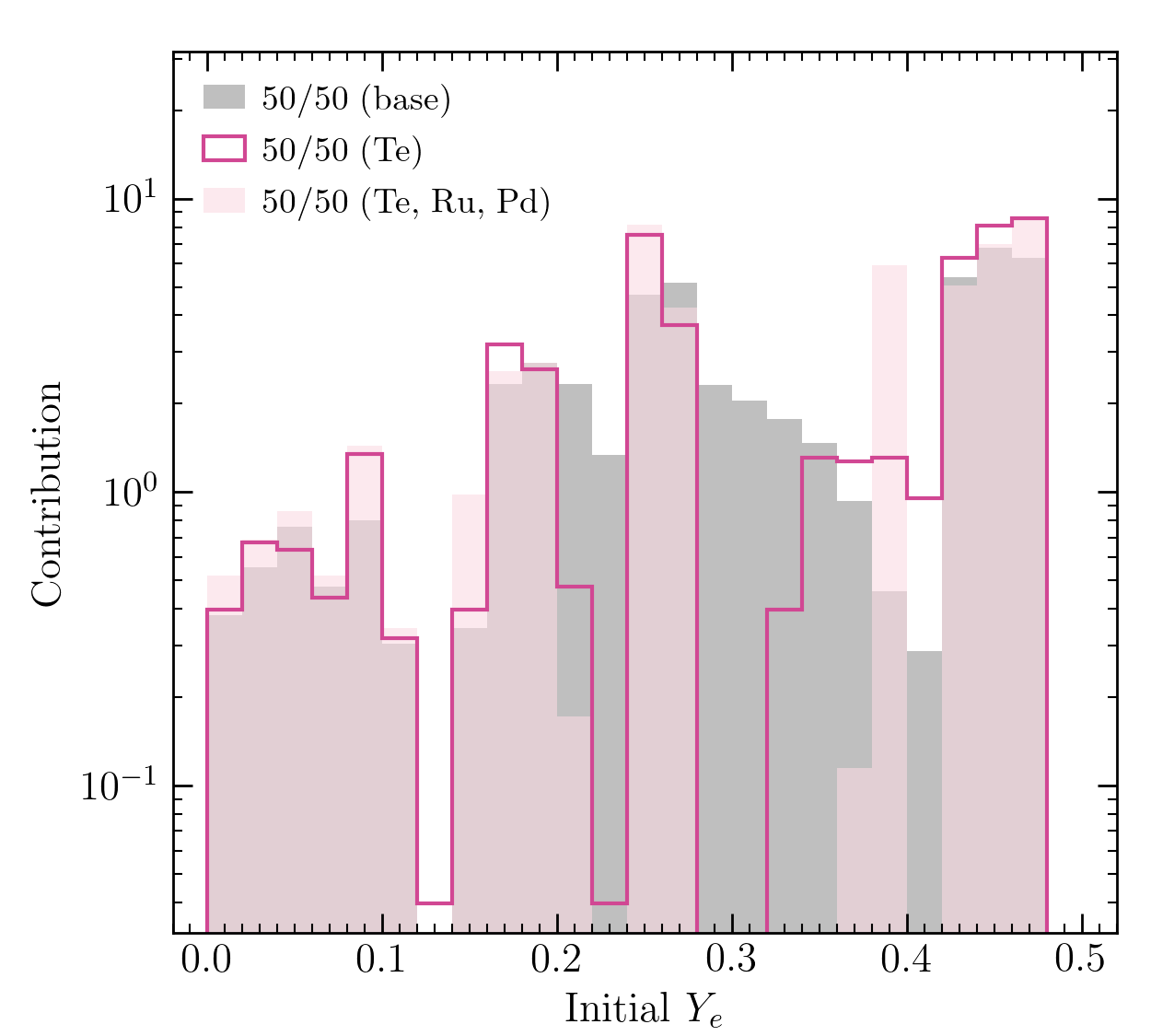}
	\caption{Combined abundance patterns (left) and $Y_e$ distributions of ejecta (right) predicted by the model when two (gray), three (teal), and five (green filled) elemental ratios are used. The individual nucleosynthesis calculations use the 50/50 fission fragment distribution for actinide-producing conditions ($Y_e\lesssim 0.25$).}
	\label{fig:s040-korobkinrh_50}
	\end{figure*}

The trend in the light elements ($38 \leq Z \leq 52$) in \mbox{HD 222925} appears to effectively require at least two \rp\ environments, supporting the second case: that one component produces up to (and effectively nothing beyond) Zr, and at least one component produces the main \rp\ pattern.
This separation in $Y_e$ suggests a decoupling of the environments that produce Zr and the main \rp\ pattern.
While possible, this decoupling also seems unlikely.
If Zr and the main \rp\ pattern are produced in decoupled sites, one might expect to find an \rpe\ star lacking in Sr.
However, there are no such \rpe\ stars known \citep{Roederer2013}
Furthermore, the variation between the light elements and the main \rp\ elements, while present, is much less than what might be expected if these elements were produced in different \rp\ sites \citep{Roederer2022b}.
Alternatively, these two sites may still inherently be from the same source (e.g., the dynamical and wind ejecta of NSMs) wherein they would always be produced, but perhaps in different relative amounts.
This explanation seems likely, provided there exists a clear separation between the $Y_e$ distribution of dynamical ejecta and that of wind ejecta (specifically at $Y_e\approx 0.34$), though current simulations do not support such a clear separation.

While either of the first two cases \emph{may} be possible, there is no strong evidence in support of either one.
It is far more likely (and scientifically interesting) that our baseline model could benefit from additional complexity, particularly as when it comes to whether the abundance pattern of \mbox{HD 222925} (especially the lighter \rp\ elements) could be reproduced by a specific combination of parameterized astrophysical conditions.
In the next sections, we search for variations on our model that are not only a good match to the \rp\ abundance pattern of \mbox{HD 222925}, but also yield a reasonable (i.e., not highly constrained) ejecta profile.
Specifically, we will increase the complexity of our ADM model to see if a different fission fragment distribution or a smooth distribution of merger ejecta can be achieved in non-parameterized trajectories.
If we still find that highly constrained ejecta is needed to reproduce the abundances derived for \mbox{HD 222925}, we are left with the last option: to challenge the nuclear data far from stability between the $N=50$ and $N=82$ closed shells.

\subsection{Fission fragments}

The fission fragment distribution employed in the initial application of the ADM model in \citet{Holmbeck2019a} uses a ``50/50" deposition scheme in which a fissioning parent nucleus produces two daughter products of identical mass and charge: $(Z,A)\rightarrow (Z/2, A/2) + (Z/2, A/2)$.
This choice is simple, as robust fission studies can vary from being symmetric about $(Z/2, A/2)$ to wide and asymmetric, even depositing in the Sr region \citep[e.g.,][]{Vassh2019,Vassh2021}.
This increase in maturity of fission studies is what motivated our decision to use FRLDM fission fragment yield for our baseline case.
However, this fission fragment distribution is broad enough to deposit in the Pd--Te region instead of being localized near the heavier side of second \emph{r}-process peak (Ba), where the 50/50 fission fragments typically deposit in low-entropy conditions.
Is the overproduction of the Pd--Te region and the disfavor of $Y_e\approx 0.34$ simply a product of broad fission yields?
For this study, we repeat the previous nucleosynthesis and ADM calculations, but instead revert back to the 50/50 fission fragment distribution to test the sensitivity of the gap at $Y_e\approx 0.34$ to fission fragment deposition.

We test both the initial baseline, two-component (Zr/Dy and Th/Dy) model as well as the more complex models which use one and three additional elements (see the previous section for details).
The results for these cases are shown in Figure \ref{fig:s040-korobkinrh_50}.
As in the previous case that uses FRLDM fission yields, the elements between Pd and Te are still over-produced in the base model (gray) line, though not as egregiously (compare especially $Z=50$ in Figures \ref{fig:s040-korobkinrh_frldm} and \ref{fig:s040-korobkinrh_50}).
In addition, a better fit to the abundance pattern is again obtained with more constraints.
However, the better fit still comes with a more constrained $Y_e$ distribution: the same effect (and at the same $Y_e$) as we saw in the baseline case that used FRLDM fission fragments.
The dearth of ejecta with $Y_e\approx 0.34$ is clearly robust to fission fragment considerations and indicates that the FRLDM fission yields are not causing the model to artificially disfavor $Y_e\approx 0.34$.
The $Y_e$ distribution for our ``canonical" astrophysical site is still highly constrained.
Therefore, fission fragment deposition is likely not the solution to obtaining a smoother distribution of ejecta $Y_e$.

\subsection{Rapidly expanding ejecta}
\label{sec:cold}

Changing the fission model or the abundance ratios are not the only ways in which the ADM model can be improved.
Perhaps it is our simple two-entropy baseline for the astrophysical trajectories that needs added complexity.
Since elemental production is sensitive to not only the initial $Y_e$ and entropy, but also the \emph{evolution} of entropy (related to temperature through the equation of state) over time, the gaps in the $Y_e$ distribution may be effectively filled by astrophysical trajectories which treat the temperature evolution differently than in our two parameterized cases.
For this test, we continue exploring alternatives to our baseline model by considering a modification to the temperature evolution in our astrophysical trajectories for the \emph{r}-process ejecta.
In addition, we have shown that adding Te as a constraint produces a better fit to the abundance pattern of \mbox{HD 222925}.
While continuing to add more abundance ratios (e.g., Ru and Pd) produces an even better fit, the fit is not so much increased as to justify the increase in computational time needed for the model to converge, and similar conclusions can be drawn about the ejecta without these additional constraints.
For the following tests, unless otherwise noted, we use Zr/Dy, Th/Dy, and Zr/Te as the default set of model constraints supplied to the ADM model.

\begin{figure}[t]
	\centering
	\includegraphics[width=\columnwidth]{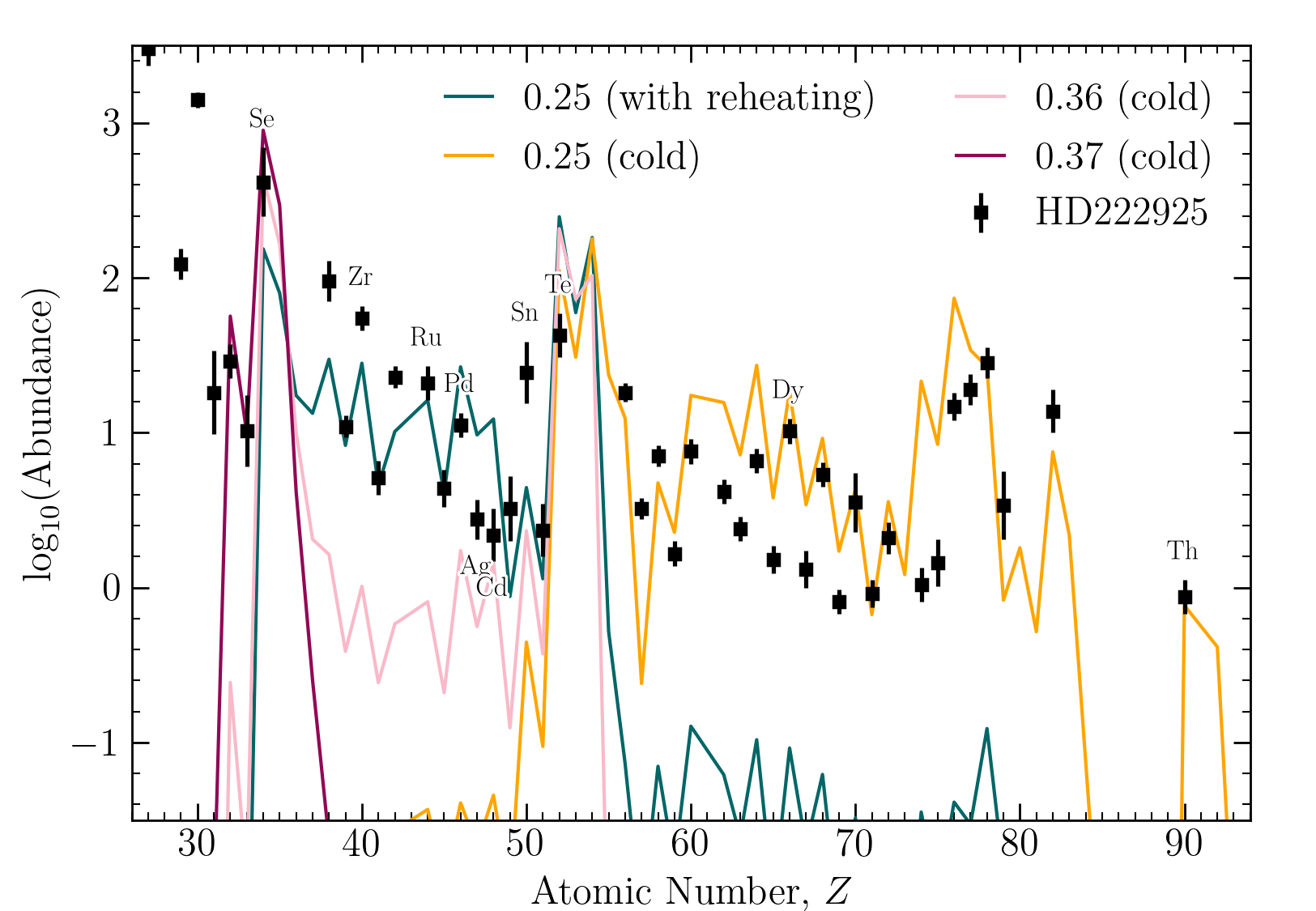}
	\caption{Abundance patterns of ``cold" (without nuclear reheating) trajectories at different initial $Y_e$ values. The teal line at $Y_e=0.25$ is an output from one of the original trajectories that uses reheating.}
	\label{fig:cold}
	\end{figure}

All of the network simulations that form the basis of our ADM models in this work include nuclear reheating; the decay of \emph{r}-process nuclei in the ejecta will add energy in the form of heat to the environment and thus affect the (temperature-dependent) reaction rates at play during the \emph{r}-process.
In our network simulations, the amount of reheating is self-consistently modified such that the amount by which the temperature changes throughout its evolution depends on the initial $Y_e$ (and evolving composition) for the same thermodynamic trajectory.
At low entropy and low-$Y_e$, the effect of reheating can govern the entire thermodynamic evolution of the trajectory and produce an abundance pattern drastically different from the case in which nuclear reheating is neglected.
For example, Figure \ref{fig:cold} shows a collection of nucleosynthesis output with and without nuclear reheating.
Even at the same $Y_e$, the inclusion of nuclear reheating can produce a completely different abundance pattern (compare the teal and orange lines at $Y_e=0.25$).
Some non-reheated cases succeed at co-producing Zr and Te, while allowing Pd, Ag, and Cd to remain extremely low (light pink line at $Y_e=0.36$).
Notably, the ``cold" trajectories produce Se at moderately low $Y_e$ ranges, some with ejecta that is almost entirely Se (dark pink line at $Y_e=0.37$).

	\begin{figure*}[t]
	\centering
	\includegraphics[height=2.6in]{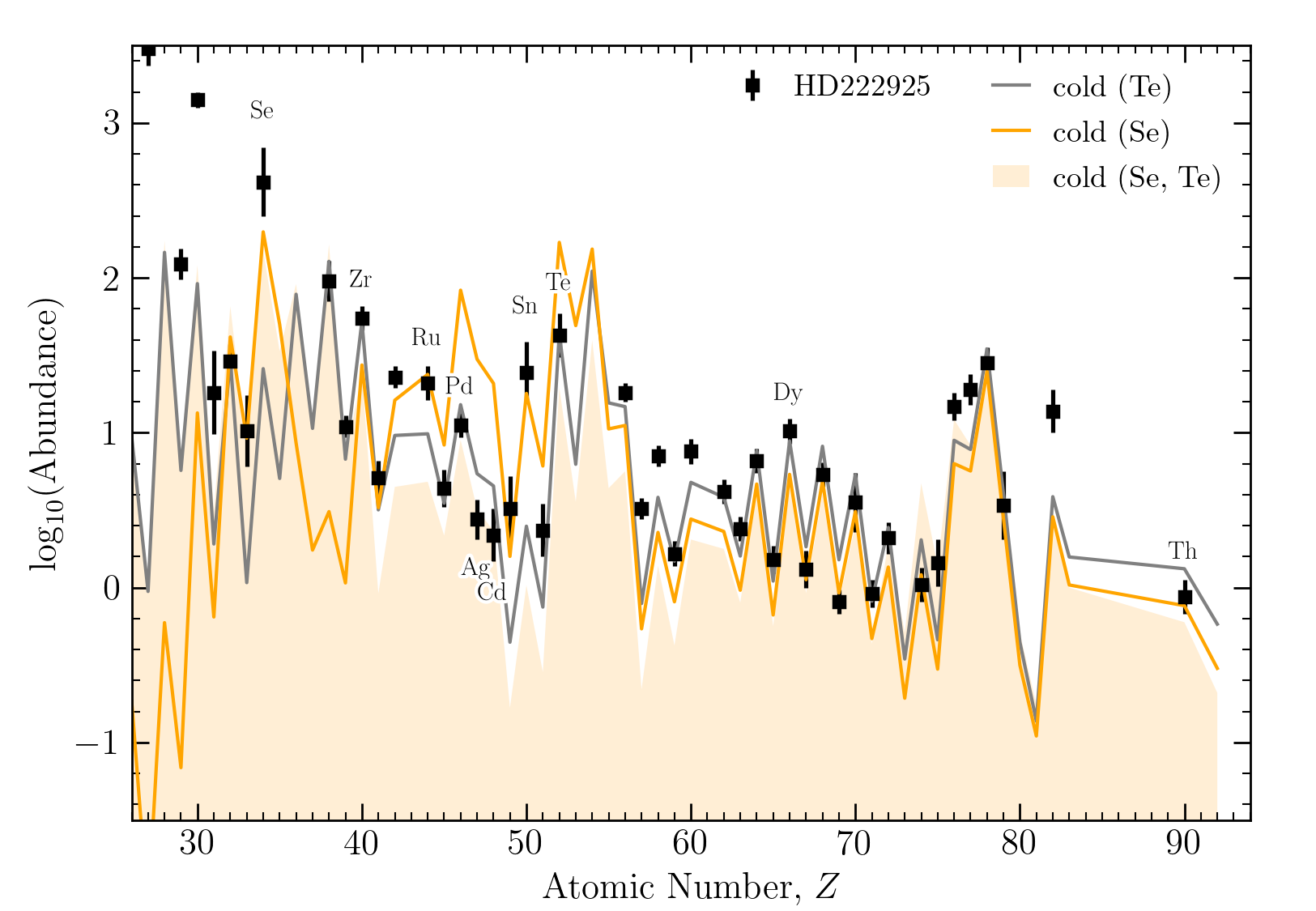}
	\includegraphics[height=2.6in]{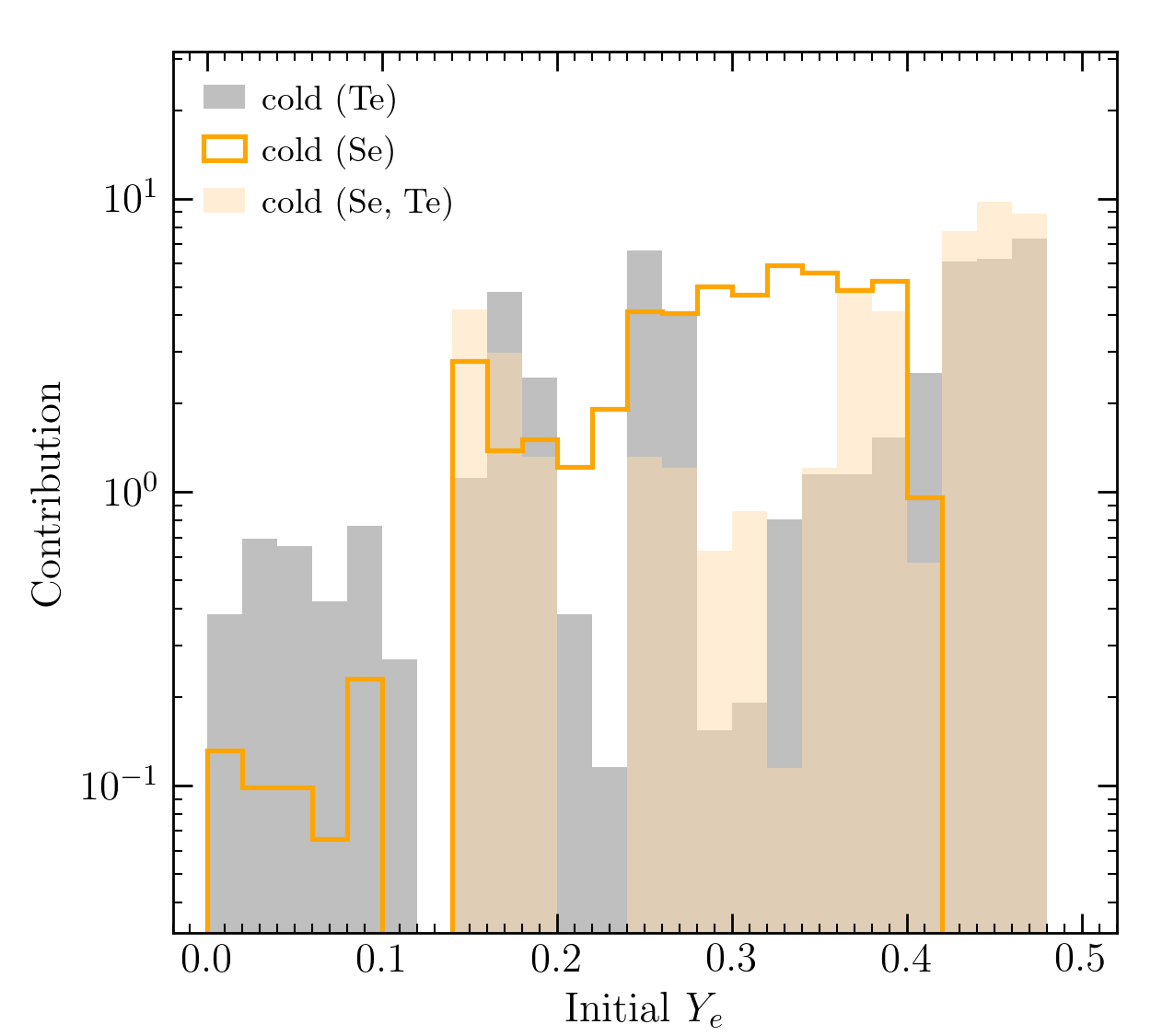}
	\caption{Combined abundance patterns (left) and $Y_e$ distributions of ejecta (right) predicted by the model when using a Te (gray) or a Se (orange line) constraint in addition to the constraints on Zr, Dy, Th (gray) and when both Te and Se are used (orange filled).}
	\label{fig:adm_cold}
	\end{figure*}

We perform a simple test wherein the low-entropy trajectories are run through PRISM without nuclear reheating included, then run the ADM model with these ``cold" trajectories at $0.35\leq Y_e\leq 0.40$ in additional to the original (nuclear-reheated) ones at $Y_e\leq 0.25$.
We still include the medium-entropy set for the model to randomly sample in range $0.25\leq Y_e\leq 0.50$.
Figure \ref{fig:adm_cold} shows the ADM results for this case (gray line) and the $Y_e$ histogram when the low-entropy, cold (i.e., non-reheated) trajectories are included.
Note that Te is used as a constraint in addition to the baseline Zr, Dy, and Th.
The $Y_e$ distribution in gray is nearly identical to the teal histogram in Figure \ref{fig:s040-korobkinrh_frldm}.
This agreement is unsurprising; the teal line uses the same elements as constraints.
In this case, the non-reheated trajectories do not need to (and barely do) contribute since a decent match can be found to the abundance pattern of \mbox{HD 222925} without them.

However, Figure \ref{fig:cold} shows that the cold trajectories are able to produce abundant Se.
Therefore, we also test how using Se as an additional constraint (in the form of the Se/Zr ratio) affects the fit and the distribution.
We test both the case where Se is used in addition to the original Zr, Dy, and Th (i.e., replacing the use of Te) and one in which both Te and Se are used.
The case with Se (and not Te) is shown by the orange line in Figure \ref{fig:adm_cold}.
Since the constraint on Te was released, the Pd to Te region is again overproduced, as we saw in the original ADM application.
However, Se is a good match to \mbox{HD 222925} and the $Y_e$ distribution for the case with Se is relatively continuous.

The case in which both Te and Se are used as constraints is shown by the orange filled region in Figure \ref{fig:adm_cold}.
There are two noteworthy features about this distribution.
First, compared to the case with only Te, the distribution appears perhaps slightly less constrained, but with more production at $Y_e\approx 0.30$ and 0.37.
At these values of initial $Y_e$, the non-reheated nucleosynthesis calculations contribute, filling in some of the gaps produced when the model is run without them.
The second noteworthy feature is that there is no material at $Y_e<0.15$ in this distribution.
In this case, the combination of heated and non-heated ejecta produces a sufficiently robust main \emph{r}-process pattern that no very low-$Y_e$ ($\lesssim 0.1$) ejecta is necessary, which further suggests that the abundance pattern of \mbox{HD 222925} does not require such extremely neutron-rich conditions.

Since the ``cold" case (with Se and Te as added constraints) leads to a somewhat less-constrained $Y_e$ distribution, we interpret this solution as a possible canonical \rp\ site: that is, that \rp\ sites at low metallicity included a significant ejecta component that did not undergo nuclear reheating.
To achieve a non-reheated case, the ejecta would have to expand very rapidly such that a sudden decrease in density effectively suppresses reheating efficiency.
For the case in Figure \ref{fig:adm_cold} to be the solution to \mbox{HD 222925}, there may need to be a correlation of $Y_e$ with ejecta velocity, with higher-$Y_e$ ejecta expanding faster than lower-$Y_e$ ejecta.
Indeed, as one example, the collapsar disk models of \citet{Miller2020} appear to show a high-velocity component at $Y_e\approx 0.35$.
The abundance pattern of \mbox{HD 222925} could be the result of an \emph{r}-process event (such as an NSM) with a low-entropy component of sufficiently high velocity to suppress nuclear reheating and produce both the high Se and Te as well as the downward trend between Zr and Te.
However, low-entropy, rapidly expanding ejecta without nuclear reheating is uncommon in NSM simulations; shock-interface ejecta, which is ejected quickly and at $Y_e\approx 0.2$--0.4 \citep{Hotokezaka2015}, typically has higher entropy than the very low-entropy case considered here.
Many unknowns still exist in our understanding of NSM ejecta.
Perhaps low-entropy, high-velocity ejecta that does not undergo nuclear reheating comprises a substantial fraction of the total NSM outflows.
Further studies on the theory side, especially in neutrino physics, are necessary to understand the complexities of merger ejecta to test whether a high-mass, non-reheated component is viable in NSM sites.
We continue to explore other possibilities and next test if a smoother entropy range can remove the gap at $Y_e\approx 0.34$.

\subsection{A smooth entropy range}

	\begin{figure*}[t]
	\centering
	\includegraphics[height=2.6in]{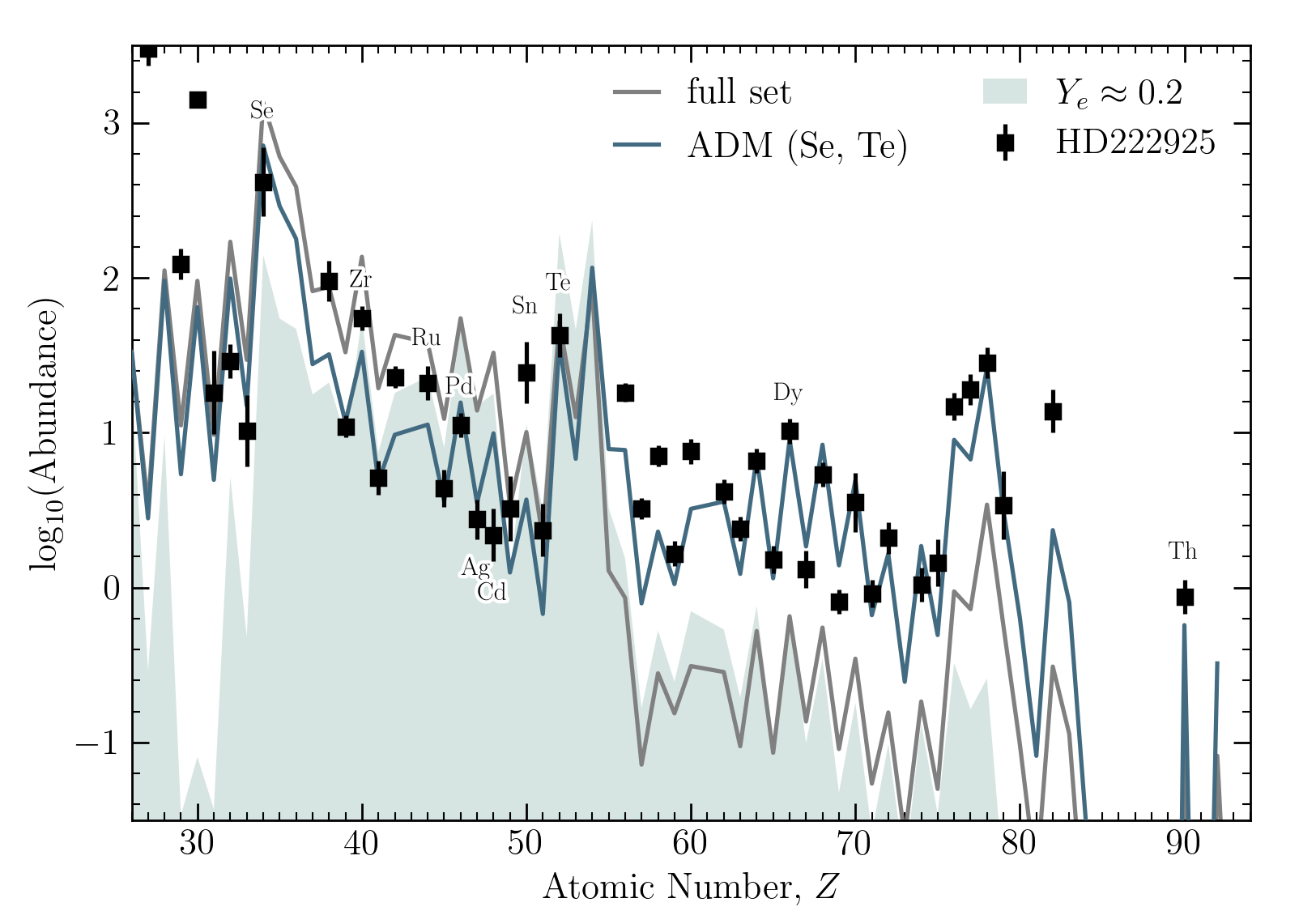}
	\includegraphics[height=2.6in]{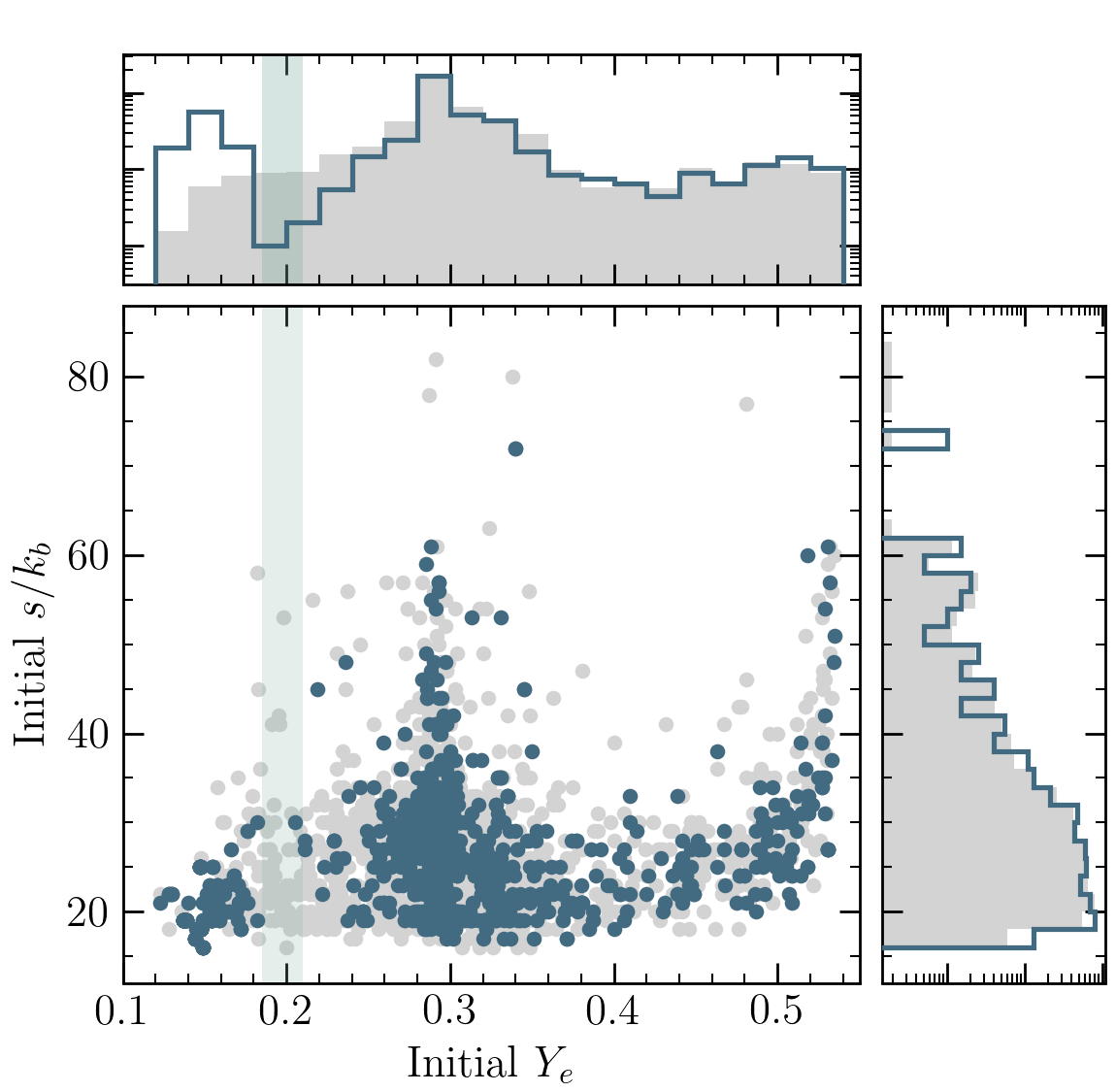}
	\caption{Combined abundance patterns (left) and $Y_e$-entropy distribution of ejecta (right) predicted by the ADM model (blue) compared to the original simulation (gray). The light-blue shaded abundance pattern in the left panel shows the average of subset of trajectories that start with a $Y_e\approx 0.2$, corresponding to the range highlighted in the right panel.}
	\label{fig:rodrigo}
	\end{figure*}

Rather than sampling from a variety of $Y_e$'s but only two entropies we instead sample from the hydrodynamic simulation output of \citet{Metzger2014}.
Although initial $Y_e$ and specific entropy of the trajectories in the simulation output do not evenly populate a regular grid, they nevertheless span a range that should be more or less in agreement with other theoretical predictions.
This sampling allows us to try not only different $Y_e$-entropy combinations, but also different dynamical timescales (essentially expansion velocities).
In addition, sampling from a simulation will also be able to capture any correlation effects between astrophysical conditions.
We take the trajectories from the 100-ms-lived hyper-massive neutron star case in \citet{Metzger2014} and randomly sample ten output abundance patterns from previously performed calculations \citep{Holmbeck2021} to test if the randomly sampled combination is a reasonable subset of the total.
For this exercise, we use both the Se and Te constraints, as in the orange shaded case in Figure \ref{fig:adm_cold}.

Figure \ref{fig:rodrigo} shows the results when, instead of randomly sampling a range of $Y_e$'s for a small handful of entropies, an entire distribution of entropy-$Y_e$ space is sampled, based on simulation data for NSMs.
The total abundance pattern and selected entropy-$Y_e$ combinations are shown from ADM (blue) compared to the original full set of trajectories (gray).
As can be seen in the left panel of Figure \ref{fig:rodrigo}, the main trends (high Se, low Pd, and Solar lanthanide-to-actinide ratios) are well-reproduced by the ADM combination.
The right panel of Figure \ref{fig:rodrigo} shows that, for the most part, the ADM-selected trajectories are indeed a reasonable subset of the total distribution.

However, there are two important differences; first, there is a preference for low $Y_e$ ($\lesssim 0.18$), dominating over the original distribution.
The preference for low $Y_e$ can be explained by comparing the blue and gray abundance patterns in the left panel of the figure.
The full set has lower overall main \emph{r}-process abundances.
For a star like \mbox{HD 222925}, which is enhanced in these elements, a higher contribution of the conditions that produce the main \emph{r}-process (low-$Y_e$, low-entropy) is preferred.
	
Secondly, there is a gap at $Y_e\approx 0.2$.
The light-blue shaded abundance pattern shows the average of all trajectories in the original sample with initial $Y_e\approx 0.2$, about 2\% of all the simulation data.
From this abundance pattern, it is clear why there is a lack of material at this $Y_e$; the Te region is overproduced, similarly to what occurred at $Y_e\approx 0.34$ in the ADM cases from the previous sections.
Because Te is used as one of the constraints for the ADM fit, it is natural that the model gravitates away from this subset of simulation output.

We have explored many astrophysical variations---and one nuclear one---and conclude that the puzzle of the $Y_e$ gap first seen in Figure \ref{fig:s040-korobkinrh_50} cannot be solved by considering a more complex model, a smooth entropy range, or a different fission fragment distribution.
(A non-heated, rapidly expanding ejecta case remains a possibility.)
In terms of astrophysics, the persistence of a highly constrained ejecta profile no matter what variations we make to the model forces us back to the original first two possibilities presented in Section~\ref{sec:base}: that the ejecta for the canonical \rp\ site is indeed highly constrained, or that more than one site is needed.
While there remains still a broad range of astrophysical parameter space to explore, we consider an alternate option outside of the realm of astrophysics: that the nuclear data responsible for shaping the abundance pattern between the first and second \emph{r}-process peaks needs to be revisited \citep{Mumpower2016}.

\section{Nuclear data revisited}
\label{sec:data}

If the abundance pattern of \mbox{HD 222925} cannot be straightforwardly explained by increasing model complexity or a continuum of astrophysical \emph{r}-process conditions (nor readily by \emph{s}-process models), then it may be that the highly constrained and possibly artificial ejecta distributions are a result of uncertain nuclear physics.
Here we examine the nuclear physics inputs used in our models, particularly the neutron-rich species between the $N=50$ and $N=82$ shell closures that shape the pattern of $44\leq Z\leq 48$ elements in \mbox{HD 222925} that fall well below that of the scaled Solar residuals.

Since $Y_e\approx 0.34$ was heavily disfavored in our initial ejecta distribution models, we investigate the specific production at this $Y_e$ with a range of nuclear mass models.
For this study, we change the separation energies, neutron-capture rates, and $\beta$-decay rates as self-consistently as possible within each mass model using state-of-the-art calculations \citep{Mumpower2015}.
(Fission and $\alpha$-decay rates are also changed for completeness, though the heaviest elements that undergo these processes are not produced in any significant amounts for these astrophysical conditions.)
In all nucleosynthesis calculations, laboratory-measured nuclear masses and decay rates were used for nuclei for which these data exist.
In other words, reaction and decay rates using different theoretical nuclear mass models only change data for which laboratory measurements do not exist, which typically means data for nuclei very far from stability.

	\begin{figure}[t]
	\centering
	\includegraphics[width=\columnwidth]{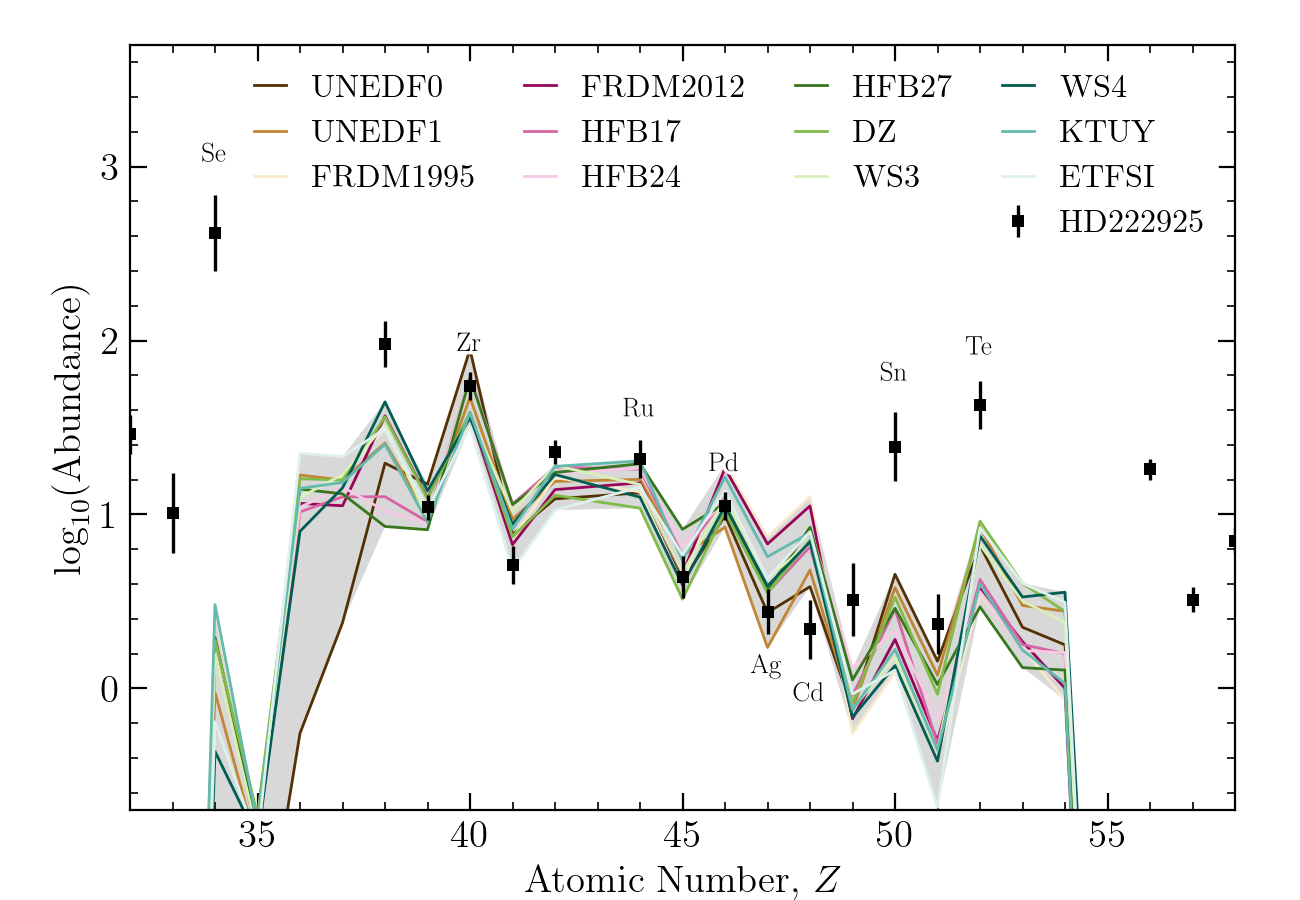}
	\caption{Abundance patterns for $Y_e=0.34$ cases with different nuclear mass models. Gray shading represents the full range spanned for each element across the mass models.}
	\label{fig:mass_models}
	\end{figure}
	
Figure~\ref{fig:mass_models} shows the results from the PRISM nuclear reaction network code for each change of nuclear mass model: UNEDF0 \citep{Kortelainen2010}, UNEDF1 \citep{Kortelainen2010}, FRDM1995 \citep{Moller1995}, HFB17 \citep{Goriely2009}, HFB24 and HFB27 \citep{Goriely2013}, DZ \citep{Duflo1995}, WS3 and WS4 \citep{Liu2011,Wang2014,Zhang2014}, KTUY \citep{Koura2005}, and ETFSI \citep{Aboussir1995}.
Our baseline case, FRDM2012, performs the worst in the $46\leq Z\leq 50$ region, overproducing Pd and Ag and underproducing Sn.
This difference highlights why the ADM model results disfavor $Y_e \approx 0.34$ (and $\approx$0.2 in the smooth-entropy case); including those conditions does not satisfy the elemental ratio constraints supplied to the model.
Note that some models like UNEDF0 fare slightly better, producing a steeper Pd-Ag-Cd trend commensurate with \mbox{HD 222925}'s elemental trend as well as a slightly higher Sn abundance.

\begin{figure}[t]
	\centering
	\includegraphics[width=\columnwidth]{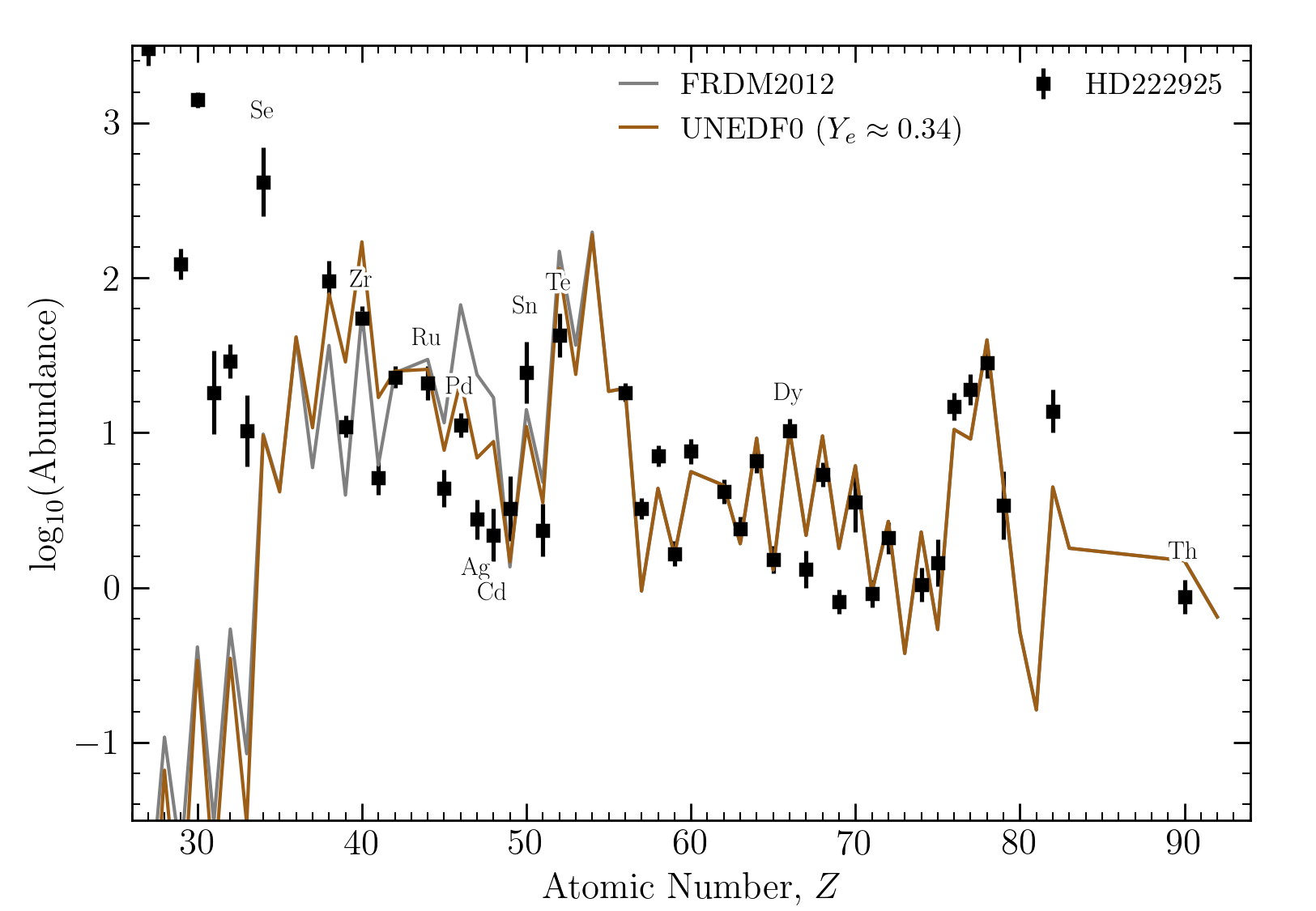}
	\caption{Gray: combined abundance pattern for the baseline case using FRDM2012 and constraints from two elemental ratios to match HD222295: Zr/Dy and Th/Dy. This pattern is the same as in Figure \ref{fig:s040-korobkinrh_frldm}. Brown: the same output results as in the baseline (gray) case, but with the original FRDM2012 output around $Y_e=0.34$ replaced by calculations that use UNEDF0.}
	\label{fig:ab_unedf0}
	\end{figure}

For a simple study, we use our baseline, two-component results (gray line in Figure \ref{fig:s040-korobkinrh_frldm}) and replace all simulations that use $0.30\leq Y_e\leq0.38$ with the UNEDF0 calculations at $Y_e\approx 0.34$.
We did not rerun the ADM model, but rather use the output from Figure \ref{fig:s040-korobkinrh_frldm} and exchanged the original (FRDM2012) output abundances with equivalent UNEDF0 abundances.
The $Y_e$ distribution, therefore, would be the same as the gray region in Figure \ref{fig:s040-korobkinrh_frldm}.
Figure \ref{fig:ab_unedf0} shows the total abundance pattern after this swap.
The overproduction originally displayed by the FRDM2012 result (and shown in \citealt{Kratz2014}) can be effectively suppressed by an exchange of nuclear mass model.
The Zr and Te abundances are slightly overproduced, but the Pd, Ag, and Cd abundances are much lower than in the FRDM2012-only case.
Without altering the astrophysics at all, changing the nuclear physics baseline leads to a case in which the main interesting features of the abundance pattern of \mbox{HD 222925} are replicated while simultaneously producing a $Y_e$ ejecta distribution that qualitatively agrees with NSM outflows from more rigorous simulations \citep[e.g.,][]{Fernandez2015,Radice2018,Fujibayashi2022}.
With a small modification to the nuclear physics, the baseline (gray) distribution could be descriptive of the canonical \rp\ site for metal-poor stars.

The dramatic effect of the underlying mass model on abundance predictions is perhaps surprising, given the experimental advances in studies of nuclei near the second \emph{r}-process peak (e.g.,\cite{Kozub2012,VanSchelt2013,Lorusso2015}).
Consequently, we did not anticipate that the choice of nuclear mass model would affect this region so significantly.
Clearly, nuclei outside of current laboratory measurements continue to drive the uncertainties on the \emph{r}-process, even at these lower-mass nuclei.
The projected reach of current and next-generation accelerators for nuclear physics (such as the Facility for Rare Isotope Beams, FRIB) are able to explore this region and contribute significantly more new data, even approaching the neutron dripline near $Z=40$.
More uncertainties remain at even higher-mass nuclei, but our investigation of \mbox{HD 222925} demonstrates that understanding the second-peak region is still a high priority for the \emph{r}-process.

\section{Summary and discussion}

\mbox{HD 222925} is second only to our Solar System for the most complete neutron-capture-element abundance pattern.
Unlike the Sun though, its low metallicity argues for a single \rp\ progenitor site that created those elements.
Since \mbox{HD 222925} represents possibly a new \rp\ template for metal-poor stars, the progenitor astrophysical site of its neutron-capture elements could be the canonical site for \rpe, metal-poor stars.
This goal was the motivation for the present study: to define and characterize the canonical \rp\ site for metal-poor stars as a function of its ejecta $Y_e$ distribution.
The model-reconstructed distribution of the $Y_e$ of the ejecta can then hint at what conditions are necessary to reproduce the elemental abundances in \mbox{HD 222925}, for example, the ratio of wind to dynamical ejecta for an NSM case.

We found that starting with our baseline two-component, two-ratio model produced a poor fit to the elemental abundance pattern of \mbox{HD 222925} between Pd and Te, where the elemental abundance pattern is uniquely complete.
We attempted to improve the fit by adding more constraints to the model.
While adding additional constraints indeed produced a better elemental fit to this region, it came at the cost of a highly---artificially---constrained $Y_e$ distribution, with a notable gap at $Y_e\approx 0.34$.
Although a highly constrained ejecta distribution could explain the origin of the elements in \mbox{HD 222925}, we explore other options to test if the gap is indeed artificial: perhaps as a result of our simplified astrophysical model or the nuclear physics basis used for the nucleosynthesis simulations.

Increasing the model complexity by considering a more robust sampling of astrophysical trajectories did not remove the gap in the $Y_e$ distribution.
Neither did changing the fission fragment distribution (with fragment deposition in the Zr to Te region) mitigate this gap.
We found a possible solution for a case in which the \rp\ material is ejected in such a way that it is not reheated by energy from the decay of its own heavy nuclei.
This case argues for a site in which the $Y_e$ of the ejecta is correlated with velocity---a situation in which the third \emph{r}-process peak could also be affected by spallation \citep{Wang2020}.
Furthermore, a site with low entropy and no nuclear reheating is exotic in terms of current NSM simulations.
For this site to be the \emph{canonical} source of \rp\ material in metal-poor stars, it would have to be fairly common and/or high-yield.
It is possible that an exotic site like this is indeed the progenitor of \mbox{HD 222925} and that this star is not canonical at all.
Alternatively, this exotic site could hint at our limited understanding of NSM ejecta; it could be that all compact binary mergers have such a low-entropy, high-velocity component (and little to no $Y_e\lesssim 0.18$ ejecta).
Either way, further studies are necessary to explore what conditions (more realistic and physics-based than our artificial study) can achieve low-entropy ejecta that also doesn't undergo nuclear reheating and if an NSM is a viable explanation for \mbox{HD 222925}.

Next, we continued to explore if the gap at $Y_e\approx 0.34$ could be an effect from the underlying nuclear physics mass model instead of requiring an exotic ejecta distribution to favor conditions that do not overproduce the Pd to Te region.
We found that at $Y_e=0.34$, our baseline nuclear mass model, FRDM2012, overproduces the Pd-Ag-Cd region compared to nearly all other mass models for the astrophysical conditions we consider.
Running a simple test wherein we replace the $Y_e\approx 0.34$ output in the baseline case that uses FRDM2012 with equivalent calculations that use UNEDF0 resulted in a better fit to the abundance pattern of \mbox{HD 222925} without the need for a highly constrained $Y_e$ ejecta distribution.
A sensitivity study on the second \emph{r}-process peak region is necessary to pinpoint which nuclei most affect the Pd, Ag, and Cd abundances in order to guide studies with next-generation accelerators such as FRIB.
For example, recent work has explored $(\alpha, xn)$ reactions and their effect on observationally derived abundance ratios between the first and second \emph{r}-process peaks \citep{Psaltis2022}.

Our main conclusions are as follows:
	\begin{itemize}
	\item A basic attempt to reproduce the abundance pattern of \mbox{HD 222925} leads to highly constrained ejecta that suggests at least two \emph{r}-process progenitor sites: one that produces light \emph{r}-process elements with $Z\lesssim 50$ and one that produces the main \emph{r}-process pattern with $Z\gtrsim 52$. This explanation is possible, but unlikely considering the observational variation between the light and heavy \emph{r}-process elements does not reflect the variation expected if these two elemental regions were not co-produced. More UV observations of \emph{r}-process-rich, metal-poor stars to determine abundances in the $46 \lesssim Z\lesssim 52$ region are necessary to investigate whether \mbox{HD 222925} is indeed typical of its class.
	\item NSM ejecta with low entropy and high velocity to suppress nuclear reheating could explain the abundance pattern of \mbox{HD 222925}. Assuming this star and its elemental origin is representative of metal-poor, \emph{r}-process-enhanced stars---and if NSMs are to be considered the primary source of heavy elements at low metallicities---models of NSMs must accommodate a high-mass component that can achieve low entropy, free from nuclear reheating.
	\item One last explanation for the elemental abundances of \mbox{HD 222925} and the failure of our models is that we are still limited by our need to use theoretical nuclear data. New laboratory measurements of nuclear masses and $\beta$-decay rates near the second \emph{r}-process peak are essential to guiding the astrophysical interpretation of our results. For example, our model shows some success with different nuclear data, possibly rendering the previous two explanations for \mbox{HD 222925} moot.
	\end{itemize}
We look forward to advances in observational astronomy, computational astrophysics, nuclear theory, and nuclear experiment to study the \emph{r}-process in light of the new, nearly complete abundance pattern of \mbox{HD 222925}.


\begin{acknowledgements}

We thank Matthew R.\ Mumpower and Trevor M.\ Sprouse for their helpful comments and PRISM support.
Support for this work was provided by NASA through the NASA Hubble Fellowship grant HST-HF2-51481.001 awarded by the Space Telescope Science Institute, which is operated by the Association of Universities for Research in Astronomy, Inc., for NASA, under contract NAS5-26555.
This work was partially enabled by the National Science Foundation (NSF) grant No.\ PHY-1430152 (Joint Institute for Nuclear Astrophysics Center for the Evolution of the Elements).
IUR acknowledges support from NSF grants AST 1815403, AST 1815767, and AST 2205847; the NASA Astrophysics Data Analysis Program (grant 80NSSC21K0627); and NASA grants HST-GO-15657 and HST-GO-15951 from the Space Telescope Science Institute.
AF acknowledges support from NSF CAREER grant AST-1255160 and NSF grant AST-1716251.
We also acknowledge support from the US Department of Energy (DOE) under contracts DE-FG02-02ER41216 (GCM), DE-FG02-95-ER40934 (RS), and LA22-ML-DE-FOA-2440 (GCM, RS).
RS and GCM acknowledge additional support from the NSF under grant No.\ PHY-2020275 (Network for Neutrinos, Astrophysics, and Symmetries) and the Heising-Simons foundation under grant No.\ 2017-228 and the Fission In R-process Elements (FIRE) topical collaboration in nuclear theory, funded by the DOE. 

\end{acknowledgements}


\bibliographystyle{aasjournal}
\bibliography{bibliography}

\end{document}